\documentclass[10pt,conference]{IEEEtran}

\usepackage[letterpaper, top=0.76in, bottom=1.05in, left=0.63in, right=0.63in]{geometry}

\IEEEoverridecommandlockouts
\usepackage{cite}
\usepackage{pgfplots}
\pgfplotsset{compat=1.18}
\usepackage{graphicx}
\usepackage{longtable}
\usepackage{enumitem} 
\usepackage{float}  
\usepackage{listings}
\usepackage{xcolor}
\usepgfplotslibrary{statistics}
\usepackage{amsmath,amssymb,amsfonts}
\usepackage{algorithmic}
\usepackage{graphicx}
\usepackage{textcomp}
\usepackage{xcolor}
\usepackage{multirow}

\usepackage{listings}
\usepackage{xcolor}

\definecolor{string}{HTML}{C41A16}
\definecolor{numb}{HTML}{1C00CF}
\definecolor{punct}{HTML}{1C1C1C}
\definecolor{delim}{HTML}{1C1C1C}
\definecolor{background}{gray}{0.97}

\lstdefinelanguage{json}{
    basicstyle=\ttfamily\small,
    numbers=none,
    showstringspaces=false,
    breaklines=true,
    frame=single,
    backgroundcolor=\color{background},
    literate=
     *{0}{{{\color{numb}0}}}{1}
      {1}{{{\color{numb}1}}}{1}
      {2}{{{\color{numb}2}}}{1}
      {3}{{{\color{numb}3}}}{1}
      {4}{{{\color{numb}4}}}{1}
      {5}{{{\color{numb}5}}}{1}
      {6}{{{\color{numb}6}}}{1}
      {7}{{{\color{numb}7}}}{1}
      {8}{{{\color{numb}8}}}{1}
      {9}{{{\color{numb}9}}}{1}
      {:}{{{\color{punct}{:}}}}{1}
      {,}{{{\color{punct}{,}}}}{1}
      {\}}{{{\color{delim}{\char`\}}}}}{1}
      {[}{{{\color{delim}{[}}}}{1}
      {]}{{{\color{delim}{]}}}}{1},
}

\lstdefinestyle{jsonstyle}{
    language=json,
    basicstyle=\ttfamily\scriptsize,
    frame=single,
    backgroundcolor=\color{background},
    breaklines=true,
    xleftmargin=2ex,    % Pulls the left side in
    xrightmargin=2ex    % Pulls the right side in
}

\def\BibTeX{{\rm B\kern-.05em{\sc i\kern-.025em b}\kern-.08em
    T\kern-.1667em\lower.7ex\hbox{E}\kern-.125emX}}

\begin{document}
\bstctlcite{IEEEexample:BSTcontrol}

\title{Voice-Driven Semantic Perception for UAV-Assisted Emergency Networks\\
\thanks{This work is co-financed by Component 5 - Capitalization and Business Innovation, integrated in the Resilience Dimension of the Recovery and Resilience Plan within the scope of the Recovery and Resilience Mechanism (MRR) of the European Union (EU), framed in the Next Generation EU, for the period 2021 - 2026, within project NEXUS, with reference 53.}}

\author{
	\IEEEauthorblockN{
	Nuno Saavedra, Pedro Ribeiro, André Coelho, Rui Campos}% <- this % stops a space
	\IEEEauthorblockA{INESC TEC and Faculdade de Engenharia, Universidade do Porto, Portugal\\
    \{nuno.m.carvalho, pedro.m.ribeiro, andre.f.coelho, rui.l.campos\}@inesctec.pt}
}

\maketitle

\begin{abstract}

Unmanned Aerial Vehicle (UAV)-assisted networks are increasingly foreseen as a promising approach for emergency response, providing rapid, flexible, and resilient communications in environments where terrestrial infrastructure is degraded or unavailable. In such scenarios, voice radio communications remain essential for first responders due to their robustness; however, their unstructured nature prevents direct integration with automated UAV-assisted network management. This paper proposes SIREN, an AI-driven framework that enables voice-driven perception for UAV-assisted networks. By integrating Automatic Speech Recognition (ASR) with Large Language Model (LLM)–based semantic extraction and Natural Language Processing (NLP) validation, SIREN converts emergency voice traffic into structured, machine-readable information, including responding units, location references, emergency severity, and Quality-of-Service (QoS) requirements. SIREN is evaluated using synthetic emergency scenarios with controlled variations in language, speaker count, background noise, and message complexity. The results demonstrate robust transcription and reliable semantic extraction across diverse operating conditions, while highlighting speaker diarization and geographic ambiguity as the main limiting factors. These findings establish the feasibility of voice-driven situational awareness for UAV-assisted networks and show a practical foundation for human-in-the-loop decision support and adaptive network management in emergency response operations.

\end{abstract}


\section{Introduction}
In emergency scenarios, first responders continue to rely on voice radio transmissions due to their robustness under harsh conditions. However, while effective for basic coordination, voice communications are inherently unstructured, limiting their integration with automated systems and constraining coordinated decision-making and situational awareness.

UAVs, acting as airborne communication relays, enable the rapid deployment of flying networks that restore connectivity in affected areas. Beyond supporting voice traffic, these networks can deliver media-rich data such as images, video, and sensor measurements, thereby enhancing situational awareness and enabling more informed operational decisions. Their inherent flexibility makes UAV-assisted networks particularly suitable for temporary and highly dynamic emergency operations, where communication demands and network topology must continuously adapt.

When integrated into UAV-assisted network management, semantic perception extracted from field communications can inform adaptive decisions such as UAV positioning, trajectory adjustment, and communication resource allocation, including the prioritization of links, bandwidth, and mission-critical traffic. In emergency operations, many of these operational needs are conveyed implicitly or explicitly through voice exchanges among first responders. Consequently, voice communications constitute a rich yet largely underutilized source of semantic perception for UAV-assisted network management.

Recent advances in Automatic Speech Recognition (ASR) and Large Language Models (LLMs)~\cite{tian2025uavs,lin2024airvista}, have enabled the transformation of unstructured voice communications into actionable information. By analyzing voice communications, these technologies can extract semantic entities such as location references, emergency severity levels, and Quality-of-Service (QoS) requirements. This semantic inference complements traditional telemetry by providing contextual awareness in scenarios where other sensing information, such as imagery, is limited, degraded, or unavailable. Although NLP-enhanced ASR has already demonstrated its value for situational awareness in domains such as air traffic control~\cite{ATCNLP}, the exploitation of voice-derived semantic perception as a direct input for adaptive UAV-assisted network management remains largely unexplored, specifically for UAV positioning and resource allocation in UAV-assisted emergency response scenarios.

\begin{figure}
    \centering
    \includegraphics[width=0.85\linewidth]{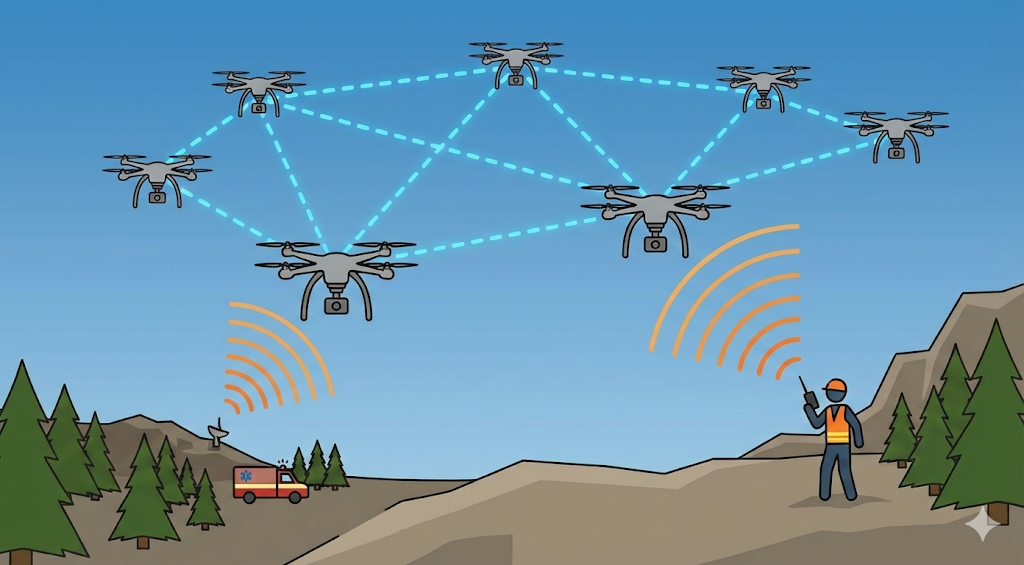}
\caption{UAV-assisted emergency network where SIREN extracts voice semantics to support ground response units.}
\label{fig:flyingNetwork}
\end{figure}

This paper proposes SIREN, an AI-driven framework designed to function as a voice-driven semantic perception layer by extracting structured information from emergency voice communications. To the best of our knowledge, SIREN is the first approach to jointly integrate ASR, LLMs, and NLP techniques, including Named Entity Recognition (NER)~\cite{nlpner} for location validation, speaker diarization for unit attribution, and sentiment analysis for emergency severity analysis, in order to interpret emergency audio streams. It is worth noting that SIREN does not perform network management or UAV deployment decisions itself; instead, it produces structured semantic outputs to support human-in-the-loop decision-making and existing network planning and control algorithms. This capability enables its direct use as a perception input for adaptive UAV-assisted network management, as illustrated in Fig.~\ref{fig:flyingNetwork}.
A preliminary version of this work was presented in~\cite{EuCNC}.

The main contributions of this work are two-fold:
\begin{enumerate}
    \item \textbf{SIREN}, a modular AI-based framework that transforms unstructured emergency voice communications into structured, machine-readable perception inputs suitable for UAV-assisted network management tasks, such as UAV positioning and communication resource allocation.

    \item \textbf{A comprehensive performance evaluation of SIREN} across five synthetic emergency communication scenarios with varying language, number of speakers, message complexity, and background noise, demonstrating the feasibility of the proposed solution while identifying key limitations, including speaker diarization errors and geographic ambiguity.

\end{enumerate}

\section{Related Work \label{sec:related-work}}
In scenarios where terrestrial infrastructure may be damaged or unavailable, UAVs can act as airborne relays and access points, where performance is strongly affected by UAV mobility and traffic demand, requiring continuous adaptation of UAV positioning and communications resources~\cite{ribeiro2024supply}. Many UAV placement solutions still rely on assumptions about user locations, traffic demand, and channel conditions~\cite{sobouti2024utilizing,ribeiro2024supply}, which may not hold in dynamic emergency environments.

Perception is a core capability in emerging UAV-assisted networks, typically driven by computer vision for detection and tracking~\cite{xiao2025vision}. While computer vision provides rich spatial information, it degrades under occlusions, illumination changes, and adverse conditions common in disasters. Audio-based perception is a promising complement to vision by reducing reliance on line-of-sight and lighting~\cite{alla2024sound}.

In parallel, LLMs are increasingly used as cognitive layers that operate on structured semantic representations to bridge perception and decision-making, enabling contextual interpretation and intent inference in UAV systems~\cite{tian2025uavs}. Representative frameworks show LLM-based spatial reasoning and multimodal interpretation when operating on structured outputs from perception modules~\cite{lin2024airvista}. Despite these advances, most LLM-enabled UAV perception frameworks remain vision-centric and focus on sensor-derived data.

The use of unstructured emergency voice communications as a semantic perception source, capturing operational intent, urgency, and contextual requirements from first responders, remains largely unexplored. Although agentic architectures for UAV-assisted networks have considered separating perception from decision-making via structured semantic representations~\cite{coelho2025a4fn}, such frameworks remain largely conceptual. The integration of voice-based operational information is proposed only as a potential future direction, lacking concrete implementations.

\section{SIREN framework \label{sec:SIREN}}
\begin{figure*}
    \centering
    \includegraphics[width=0.78\textwidth]{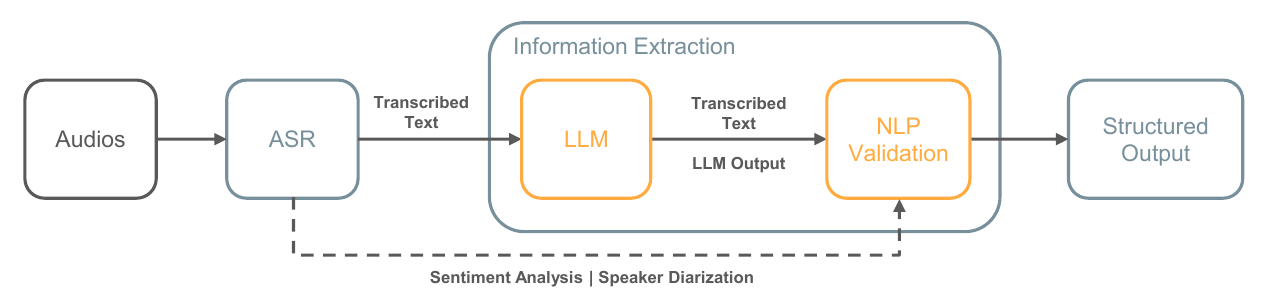}
    \caption{SIREN pipeline, comprising ASR, Information Extraction, integrating LLM inference with NLP validation, and generation of structured outputs.}
    \label{fig:pipeline}
\end{figure*}

This section introduces SIREN, an AI-driven framework that converts unstructured emergency voice communications into structured, machine-readable semantic representations. We describe the end-to-end pipeline (cf. Fig.~\ref{fig:pipeline}) from raw audio to structured outputs, designed for integration with UAV-assisted network management functions such as UAV positioning and communications resource allocation.

\subsection{System design}

Each stage of the pipeline is described in the following.

\subsubsection{Automatic Speech Recognition}
The ASR stage performs speech-to-text conversion and is designed to accommodate different approaches, depending on the operational context. The pipeline supports the integration of lightweight local models for offline processing in connectivity-constrained environments, as well as cloud-based APIs that allow for high-accuracy transcription, especially under noisy conditions, and enable advanced features such as speaker diarization and sentiment analysis. 

\subsubsection{Information extraction}
The information extraction stage processes the transcribed text through a modular LLM component, which can be deployed either locally or via cloud-based services. To ensure reliability, the system employs schema-constrained prompting to define specific output fields, such as \textit{location}, \textit{units}, \textit{emergency\_level}, and \textit{QoS expectations}. The LLM output is validated against this schema to mitigate hallucinations and to ensure interoperability with UAV control and network management systems (cf.\ Listing~\ref{lst:json_schema}).

To further enhance reliability, the extracted information is cross-validated using complementary NLP techniques. NER is applied to verify geographic entities, and, when a cloud-based API is available, speaker diarization and sentiment analysis may be employed to improve unit attribution and to refine emergency severity estimation, respectively.

\subsubsection{Output information}

The SIREN framework produces a structured representation (e.g., a JSON object) designed for integration with state-of-the-art solutions for network management, such as UAV positioning, network planning, and communications resource allocation. Listing~\ref{lst:json_schema} provides a reference example of this representation and its main semantic categories. This structured output may also serve as support for human-in-the-loop decision-making.

The structured output includes three main semantic categories. First, the \textit{locations} field aggregates geographic references mentioned in the communication (e.g., street names or landmarks). Second, \textit{the emergency\_level} field indicates the severity of the situation (e.g., critical, minor). These emergency level categories are predefined in the prompt, and the model selects one based solely on the content of the communication. Third, the \textit{units} structure describes the active responders and their associated attributes. It is worth mentioning that only information explicitly stated in the communication is included; no additional inference or geocoding-based enrichment is performed.

\begin{figure}[htbp]
\centering
\begin{lstlisting}[style=jsonstyle]
{
  "locations": [
    "Stow Lake",
    "Gold Star Mother's Rock"
  ],
  "emergency_level": "Critical",
  "units": [
    {
      "name": "Unit Alpha",
      "location": "Gold Star Mother's Rock",
      "video_support": {
        "needed": true,
        "issue": "Poor signal; need uplink for fire assessment.",
        "priority": "high",
        "requirements": "10 Mbit/s"
      },
      "times_intervened": 6
    }
  ]
}
\end{lstlisting}
\caption{Example of structured semantic output generated by SIREN from emergency voice communications.}
\label{lst:json_schema}
\end{figure}

For each identified unit, SIREN associates the reported location reference with the extracted locations and populates the \textit{times\_intervened} field, which reflects the number of interactions in the communication and indicates operational relevance. The \textit{video\_support} and \textit{image\_support} fields capture QoS expectations, including the requested support type, a brief justification, a priority level, and an indicative throughput requirement (in $\mathrm{Mbit/s}$). These throughput values are intended as expectations rather than precise measurements. Requiring explicit justifications further acts as a grounding mechanism, improving transparency, reducing ambiguity in support requests, and mitigating the risk of hallucinated requirements.

\begin{figure}
    \centering
    \includegraphics[width=0.85\linewidth]{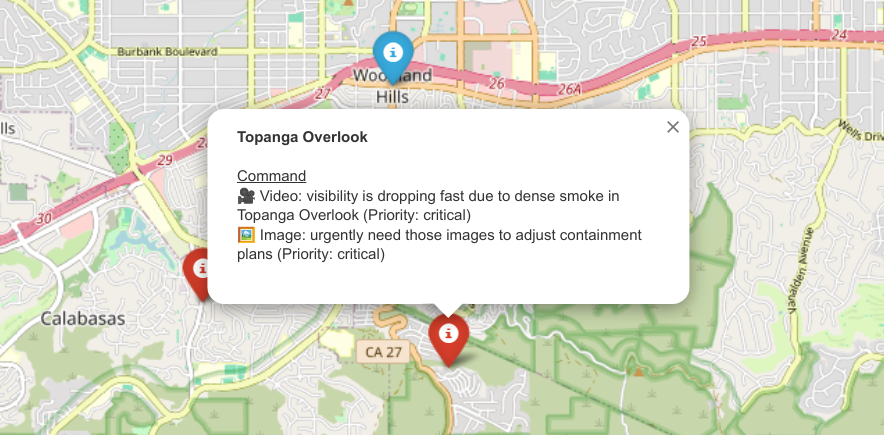}
    \caption{Interactive map interface depicting a geo-referenced unit and its specific communication requirements extracted by SIREN.}
    \label{fig:mapSample}
\end{figure}

\subsection{System implementation}
In its current implementation, SIREN’s pipeline relies on a local instance of OpenAI Whisper~\cite{whisper} for offline ASR or on the Assembly API~\cite{assemblyai2025docs}, a cloud-based ASR API. Information extraction is handled by the LLaMA~3.2 model~\cite{touvron2024llama3} deployed locally via the Ollama framework~\cite{ollama2024}, which provides the inference engine to identify location references, unit identifiers, emergency levels, and QoS expectations. To enforce the design's requirement for structured data, a strict JSON schema is applied at the prompt level. 

Given the probabilistic nature of the LLM outputs, they are audited by a validation layer comprising three deterministic NLP-based modules. First, \textbf{Named Entity Recognition (NER)} via the SpaCy library~\cite{spacy} cross-checks geographic entities extracted by the LLM. Second, \textbf{speaker diarization} results from the Assembly API are cross-referenced with extracted units to flag potential hallucinations. Third, \textbf{sentiment analysis} provided by the API calibrates the emergency severity level (e.g., escalating a "moderate" classification to "severe" if distress is detected).

Finally, validated location entities are mapped to geographic coordinates using Geopy~\cite{geopy} and rendered through an interactive web-based interface using Folium~\cite{folium}. As depicted in Fig.~\ref{fig:mapSample}, this interactive map displays the identified units and their specific communications requirements, providing operators with quick situational awareness. While the structured output is designed to support autonomous network management, this visualization ensures that human operators can monitor and validate the semantic context of voice communications.

\section{Performance evaluation and results \label{sec:performance-evaluation}}

This section presents the performance evaluation of SIREN in terms of transcription accuracy, output quality, and execution time. The goal is to assess the feasibility and behavior of SIREN under diverse operating conditions.

\subsection{Audio dataset~\label{sec:dataset}}

Authentic emergency communication datasets are scarce due to privacy, legal, and sensitivity constraints. A reference state-of-the-art dataset, RescueSpeech \cite{sagar2023rescuespeechgermancorpusspeech}, contains annotated German search-and-rescue (SAR) recordings; however, it lacks continuous, multi-speaker conversational structure, making it unsuitable for the objectives of this work. Consequently, we generated a synthetic dataset in two stages: LLMs created coherent, multi-speaker emergency dialogues, which were then converted into speech using the ElevenLabs text-to-speech (TTS) system \footnote{https://github.com/Slyfenon/SIREN-audio\_files}. Multiple distinct voices were employed to emulate different first responders for increased realism. Additionally, noisy variants of the original audio were generated using Python post-processing techniques to assess robustness under degraded acoustic conditions.

The dataset is organized into five scenarios grouped into three levels of complexity, designed to assess the SIREN framework across linguistic and operational dimensions:

\begin{itemize}
    \item \textbf{Low-complexity scenario}
    \begin{itemize}
        \item \textbf{Scenario~1:} It is the simplest scenario. It consists of short, well-structured emergency dialogues in English, involving four distinct speakers and three location references. Both clean and noisy audio versions are included. Duration: 1~min~10~s.
    \end{itemize}

    \item \textbf{Medium-complexity scenarios}
    \begin{itemize}
        \item \textbf{Scenario~2:} This scenario increases difficulty by introducing longer dialogues and richer operational context. It maintains four speakers but includes a greater number of location references. Duration: 1~min~28~s.
        \item \textbf{Scenario~3:} Closely mirrors the structure of Scenario~2 to validate consistency. It features similar linguistic patterns, speaker and location counts as Scenario~2. Duration: 1~min~30~s.
    \end{itemize}

    \item \textbf{High-complexity scenarios}
    \begin{itemize}
        \item \textbf{Scenario~4:} This scenario introduces audio in Portuguese language, referencing geographical locations in Portugal. It contains four speakers, language-specific characteristics, such as pronunciation differences, syntactic variation, and ambiguous place names. These introduce further challenges, particularly for models optimized for English. Duration: 1~min~54~s.
        \item \textbf{Scenario~5:} The most complex scenario. It includes six distinct speakers, an increased number of locations, and dialogue that is less explicit regarding speaker identification and QoS expectations. Duration: 1~min~41~s.
    \end{itemize}
\end{itemize}

\subsection{Hardware specification }
All experiments were conducted on a single machine with the following specifications: Ubuntu~22.04 operating system, an AMD Ryzen~5~5600H processor (12 threads at 3.3~GHz), and an NVIDIA GeForce RTX~3050 laptop GPU with 4~GB of VRAM. To ensure measurement accuracy and consistency, no significant background processes were active during testing.

\subsection{Transcription quality assessment}

Transcription quality was evaluated using the Word Error Rate (WER), a standard performance metric for ASR systems, as defined in Eq.~\ref{eq:wer}, where $S$, $D$, and $I$ represent the number of substitutions, deletions, and insertions, respectively, and $N$ is the number of words in the reference transcript, as defined in~\cite{zechner2000wer}.

\begin{equation}
\mathrm{WER} = \frac{S + D + I}{N},
\label{eq:wer}
\end{equation}

Table~\ref{tab:merged_wer} compares the performance of the Assembly API-based and local Whisper models under clean and noisy audio conditions. Results are reported as WER (\%) with the corresponding substitution, deletion, and insertion counts shown in parentheses as (S/D/I). In clean audio conditions, both models achieve relatively low WER values; however, the API-based model consistently outperforms the local model, primarily due to fewer substitution errors.

Under noisy conditions, WER increases for both models; however, the degradation is substantially greater for the local Whisper model. This effect is most evident in Scenario~2, where the local Whisper WER rises from 15.31\% to 39.80\%, while the API-based model shows a more moderate increase from 12.76\% to 17.86\% (mean over 10 iterations).
The observed differences can be justified by model size. The local Whisper implementation uses a smaller model variant due to hardware constraints, limiting its ability to generalize under noisy conditions. These results indicate that transcription quality is primarily influenced by model capacity and training data. As such, the external API-based model is better suited for deployment in settings with variable or degraded audio quality, as it offers higher transcription accuracy and robustness.

\begin{table}[]
\caption{WER and Error Breakdown (Clean vs Noisy Audio)}
\centering
\resizebox{\columnwidth}{!}{
\begin{tabular}{|l|cc|cc|}
\hline
\textbf{Scenario} 
& \multicolumn{2}{c|}{\textbf{Assembly API}} 
& \multicolumn{2}{c|}{\textbf{Local Whisper}} \\
\cline{2-5}
& \textbf{Clean} & \textbf{Noisy} & \textbf{Clean} & \textbf{Noisy} \\
\hline
1 & 7.84 (6/0/6)    & 8.50 (7/0/6)    & 12.42 (14/0/5) & 14.38 (15/2/5) \\
2 & 12.76 (15/0/10) & 17.86 (22/3/10) & 15.31 (19/1/10) & 39.80 (52/10/16) \\
3 & 10.64 (13/2/10) & 10.64 (13/2/10) & 10.64 (9/8/8)   & 13.19 (20/2/9) \\
4 & 6.25 (9/0/6)    & 7.50 (12/0/6)   & 10.42 (15/0/10) & 12.08 (19/0/10) \\
5 & 18.67 (23/3/16) & 16.89 (22/1/15) & 17.78 (22/0/18) & 18.67 (24/1/17) \\
\hline
\textbf{Total} 
& \textbf{11.34 (66/5/48)} 
& \textbf{12.30 (76/6/47)} 
& \textbf{13.25 (79/9/51)} 
& \textbf{19.26 (130/15/57)} \\
\hline
\end{tabular}
}
\label{tab:merged_wer}
\end{table}

\subsection{Output quality assessment}
Output quality was evaluated across the same five scenarios that comprise the created dataset, as described in Section~\ref{sec:dataset}. Each scenario included five independent runs and featured both clean and noisy audio conditions. All runs were processed using the external transcription API, as the transcription quality assessment indicated higher transcription accuracy and stability compared to the local Whisper model.

\subsubsection{Evaluation metrics}

We assessed output quality along four dimensions: i) location identification; ii) unit extraction; iii) speaker count estimation; and iv) QoS expectation extraction.

\begin{itemize}
    \item \textbf{Location identification:} This measures the ability to extract geographic references explicitly mentioned in the audio, such as street names or landmarks, which are central to situational awareness. Performance relies on the accuracy of the transcription. Since geocoding is performed by an external service, any geocoding errors are not attributed to SIREN.

    \item \textbf{Unit extraction:} This measures the ability to identify operational units and associate them with their corresponding location references. Accurate unit identification is crucial for emergency coordination, as mislabeling a unit or incorrectly attributing its location can lead to suboptimal resource allocation.
    
    \item \textbf{Speaker count estimation:} This measures the diarization component's ability to estimate the number of distinct speakers present in the audio, even in cases with similar vocal characteristics. Errors in speaker differentiation can reduce the interpretability of the extracted semantics.
    
    \item \textbf{QoS expectation extraction:} This assesses the ability to infer operational intent related to communications needs, such as multimedia support or additional resources, based on semantic understanding.
\end{itemize}

All dimensions were assessed through manual inspection of the synthetic ground-truth dialogues. Table \ref{tab:merged_quality} reports success rates across the five scenarios under clean and noisy conditions, where each value represents the proportion of successful runs (out of five) for a given scenario and metric.

\subsubsection{Results}
Results are reported on a per-scenario basis to analyze how increasing linguistic and operational complexity affects the performance of the SIREN framework.

\paragraph{Scenario~1}
In the baseline scenario, SIREN achieved 100\% success in location identification, unit extraction, and speaker count estimation under both clean and noisy conditions (Table~\ref{tab:merged_quality}). The primary limitation lay in QoS expectation extraction, which achieved only 40\% success. This indicates that SIREN misinterpreted the relationships between units and requested actions. Therefore, QoS extraction was more sensitive to semantic ambiguity than entity extraction, even in dialogues of low complexity.

\paragraph{Scenario~2}
Location and unit extraction remained high, but speaker count estimation dropped to 0\%. Two of the four TTS voices shared similar timbre, causing the diarization model to cluster them. This occurred regardless of noise, isolating voice similarity as the dominant failure factor.

\paragraph{Scenario~3}
It resulted in stable performance, achieving 100\% success in both unit extraction and QoS expectation extraction. However, speaker count estimation recorded 0\% success due to recurring similarities among synthetic voices, as noted in Scenario~2. A minor mismatch was observed in the geocoding stage, where one extracted location was resolved to an incorrect coordinate by the external service; nevertheless, the textual entity was correctly extracted by SIREN. Aside from this external dependency, the results remained consistent under both clean and noisy conditions.

\paragraph{Scenario~4}
SIREN achieved 100\% success across all evaluated dimensions under both clean and noisy conditions. While occasional incorrect coordinate resolutions occurred due to the external geocoding service when place names had multiple plausible matches (an aspect common in Portugal), this does not indicate an extraction failure. Under noisy conditions, SIREN also demonstrates successful semantic extraction.

\paragraph{Scenario~5}
The highest-complexity scenario revealed limitations primarily related to external ambiguity and diarization. In this scenario, no run achieved complete location identification, with the number of correctly identified locations varying across runs. Similar to previous scenarios, the geocoding service often resolved them to incorrect coordinates when name ambiguity existed. Additionally, speaker count estimation degraded, as several of the six TTS-generated voices had similar acoustic characteristics. In contrast, unit extraction remained robust (three out of five), and QoS expectation extraction was fully successful (five out of five). These results highlighted speaker diarization and geocoding ambiguity as the primary bottlenecks in high-complexity audio conditions.

\begin{table}[]
\caption{Output Quality Across Scenarios (Clean vs Noisy Audio)}
\centering
\resizebox{\columnwidth}{!}{
\begin{tabular}{|c|cc|cc|cc|cc|}
\hline
\textbf{Scenario} 
& \multicolumn{2}{c|}{\textbf{Location}} 
& \multicolumn{2}{c|}{\textbf{Unit}} 
& \multicolumn{2}{c|}{\textbf{Speaker Count}} 
& \multicolumn{2}{c|}{\textbf{QoS}} \\
\cline{2-9}
& \textbf{Clean} & \textbf{Noisy} 
& \textbf{Clean} & \textbf{Noisy} 
& \textbf{Clean} & \textbf{Noisy} 
& \textbf{Clean} & \textbf{Noisy} \\
\hline
1 & 100\% & 100\% & 100\% & 100\% & 100\% & 100\% & 40\%  & 40\% \\
2 & 100\% & 80\%  & 100\% & 80\%  & 0\%   & 0\%   & 100\% & 100\% \\
3 & 100\% & 100\% & 100\% & 100\% & 0\%   & 0\%   & 100\% & 100\% \\
4 & 100\% & 100\% & 100\% & 100\% & 100\% & 100\% & 100\% & 100\% \\
5 & 0\%   & 0\%   & 60\%  & 60\%  & 0\%   & 0\%   & 100\% & 100\% \\
\hline
\end{tabular}
}
\label{tab:merged_quality}
\end{table}

\subsection{Execution time assessment}

SIREN’s execution time was averaged across ten runs per scenario, with total latency split into external API transcription and local LLM semantic analysis. As shown in Table~\ref{tab:scenario_timing}, execution time scales predictably with scenario complexity. Scenario 1 required the shortest processing time due to limited audio duration and simple semantics. However, as linguistic complexity increased, multiple speakers were introduced (Scenarios 2, 3, and 5), or non-English audio was processed (Scenario 4), local LLM inference became the dominant computational bottleneck. External API transcription latency fluctuates with network conditions; the pipeline's overall execution time is primarily constrained by the local hardware's LLM processing capacity.

\begin{table}
\caption{Average execution time per scenario and SIREN's pipeline stage.}
\centering
\begin{tabular}{|c|c|c|c|}
\hline
\textbf{Scenario} & \textbf{Transcription (s)} & \textbf{LLM (s)} & \textbf{Total (s)} \\
\hline
1  & 10.59 & 14.20 & 24.87 \\
2  & 29.48 & 22.18 & 51.67 \\
3  & 23.18 & 23.48 & 46.67 \\
4 & 13.95 & 27.45 & 41.40 \\
5  & 15.55 & 48.92 & 64.47 \\
\hline
\textbf{Mean} & \textbf{18.55} & \textbf{27.25} & \textbf{45.82} \\
\hline
\end{tabular}
\label{tab:scenario_timing}
\end{table}

\section{Discussion \label{sec:discussion}}
The evaluation results indicate that using LLMs to process voice communications and extract perception-level information relevant to UAV-assisted network management is feasible, while also highlighting limitations that affect end-to-end reliability. A primary design trade-off concerns the choice of ASR backend. The comparison between the local Whisper model and the Assembly API-based transcription shows that, while the local configuration enables offline operation, its error rate increases significantly under noise (from approximately 15\% to nearly 40\%), whereas the API-based model presents a more moderate degradation. These results indicate that offline deployment may require more capable hardware, to support larger and more robust ASR models, or alternative noise-robust front-ends, to maintain comparable transcription quality in degraded acoustic conditions.

SIREN is able to extract location references and QoS expectations from the transcript; however, mapping extracted entities to coordinates introduces errors when place names are ambiguous. In particular, the external geocoding service may resolve common street names to incorrect places. This indicates that the main limitation lies in geographic disambiguation rather than in SIREN’s entity extraction. Such findings motivate context-aware disambiguation strategies (e.g., bounding regions, operational maps) when converting extracted toponyms into actionable coordinates.

In the conducted experiments, SIREN shows limitations in speaker differentiation in scenarios with acoustically similar voice, this effect is exacerbated by synthetic text-to-speech voices with similar timbre and prosody, which can lead the diarization model to cluster distinct speakers. While human speakers may exhibit greater variability, real-world radio communications often involve compression and channel artifacts that attenuate speaker-specific cues. Consequently, diarization may be a potential source of attribution errors, especially in multi-speaker settings or under channel degradation.

Geographic ambiguity can be reduced when communications reference distinctive landmarks (e.g., \textit{Fort Point National Historic Site}) instead of generic street names that allow for multiple interpretations. Similarly, the extraction of QoS expectations is more reliable when requirements are explicitly expressed. Finally, diarization errors are mitigated in communications with clearer turn-taking and reduced speaker overlap, which improves unit attribution (e.g., \textit{Unit Alpha} and \textit{Unit Charlie}).

\section{Conclusions \label{sec:conclusions}}

This paper proposed SIREN and demonstrated that it can convert unstructured emergency voice communications into structured, machine-readable information, achieving high transcription accuracy and reliably extracting locations and responding units in both clean and noisy conditions. The results confirm the feasibility of voice-driven semantic extraction as a decision-support mechanism for UAV-assisted emergency networks. The performance evaluation demonstrated that SIREN offers near-real-time situational awareness to network operators and serves as a viable proof of concept for decision support. 
Future work will focus on enhancing geographic disambiguation by utilizing constrained search spaces and domain-specific location dictionaries. Additionally, we aim to validate the framework with authentic emergency recordings to better capture real-world radio phenomena, such as interference, signal dropouts, and ambient noise. 

\bibliographystyle{IEEEtran}
\bibliography{refs}

@IEEEtranBSTCTL{IEEEexample:BSTcontrol,
  CTLuse_forced_etal       = "yes",
  CTLmax_names_forced_etal = "3",
  CTLnames_show_etal       = "1"
}

@inproceedings{EuCNC,
    title        = {{Preliminary Approach to Voice-Driven Semantic Perception for UAV-Assisted Emergency Systems}},
	author       = {Saavedra, Nuno and Ribeiro, Pedro and Coelho, André and Campos, Rui},
	year         = 2026,
	booktitle    = {2026 Joint European Conference on Networks and Communications \& 6G Summit (EuCNC/6G Summit) (To appear)},
	volume       = {},
	number       = {},
	pages        = {},
}

@misc{sagar2023rescuespeechgermancorpusspeech,
	title        = {{RescueSpeech: A German Corpus for Speech Recognition in Search and Rescue Domain}},
	author       = {Sangeet Sagar and Mirco Ravanelli and Bernd Kiefer and Ivana Kruijff Korbayova and Josef van Genabith},
	year         = 2023,
	url          = {https://arxiv.org/abs/2306.04054},
	eprint       = {2306.04054},
	archiveprefix = {arXiv},
	primaryclass = {eess.AS}
}

@misc{whisper,
	title        = {{Whisper: OpenAI’s Speech Recognition Model}},
	author       = {{OpenAI}},
	year         = 2022,
	note         = "{Accessed: Jan. 16, 2026}",
	howpublished = {\url{https://github.com/openai/whisper}}
}

@misc{assemblyai2025docs,
	title        = {{AssemblyAI Documentation}},
	author       = {AssemblyAI},
	year         = 2025,
	note         = "{Accessed: Jan. 16, 2026}",
	howpublished = {\url{https://www.assemblyai.com/docs}}
}

@article{zechner2000wer,
	title        = {{Minimizing Word Error Rate in Textual Summaries of Spoken Language}},
	author       = {Zechner, Klaus and Waibel, Alex},
	year         = 2000,
	journal      = {Language Technologies Institute, Carnegie Mellon University},
	address      = {5000 Forbes Avenue, Pittsburgh, PA 15213, USA},
	note         = {\url{https://aclanthology.org/A00-2025.pdf}}
}

@misc{folium,
	title        = {{Folium}},
	author       = {Folium},
	note         = "{Accessed: Jan. 16, 2026}",
	url          = {https://python-visualization.github.io/folium/latest/}
}

@misc{geopy,
	title        = {{GeoPy: Geocoding library for Python}},
	author       = {GeoPy},
	note         = {{Accessed: 2026-1-16}},
	url          = {https://geopy.readthedocs.io/en/stable/}
}

@misc{touvron2024llama3,
	title        = {{LLaMA 3: Open Foundation and Instruction Models}},
	author       = {Touvron, Hugo and Lavril, Thibaut and Izacard, Gautier and Martinet, Xavier and others},
	year         = 2024,
	note         = {Meta AI},
	url          = {https://ai.meta.com/llama}
}

@article{nlpner,
  title   = "Named Entity Recognition ({NER}) in {NLP} Techniques, Tools
             Accuracy and Performance",
  author  = "Naseer, S and Ghafoor, M and Alvi, S and Kiran, A and Shafique Ur
             Rahmand, Ghulam and Murtazae, G",
  journal = "Pakistan Journal of Multidisciplinary Research",
  volume  =  2,
  number  =  2,
  pages   = "293--308",
  month   =  jan,
  year    =  2022
}

@misc{ollama2024,
	title        = {{Ollama: Get up and running with large language models.}},
	author       = {Ollama},
	year         = 2024,
	note         = "{Accessed: Jan. 16, 2026}",
	howpublished = {\url{https://ollama.com}}
}

@article{ATCNLP,
	title        = {{Intelligent air traffic control using NLP-enhanced speech recognition and natural language generation}},
	author       = {Sarhan, Amany and Fathy, Rawda and Ali, Hesham},
	year         = 2025,
	month        = {07},
	journal      = {Journal of Electrical Systems and Information Technology},
	volume       = 12,
	pages        = {},
	doi          = {10.1186/s43067-025-00234-9}
}

@misc{spacy,
	title        = {{{spaCy}: Industrial-strength Natural Language Processing in Python}},
	author       = {Honnibal, Matthew and Montani, Ines and Van Landeghem, Sofie and Boyd, Adriane},
	year         = 2020,
	publisher    = {Zenodo},
	doi          = {10.5281/zenodo.1212303},
	url          = {https://doi.org/10.5281/zenodo.1212303}
}

@article{sobouti2024utilizing,
	title        = {{Utilizing UAVs in Wireless Networks: Advantages, Challenges, Objectives, and Solution Methods}},
	author       = {Sobouti, Mohammad Javad and Mohajerzadeh, Amirhossein and Adarbah, Haitham Y. and Rahimi, Zahra and Ahmadi, Hamed},
	year         = 2024,
	journal      = {Vehicles},
	volume       = 6,
	number       = 4,
	pages        = {1769--1800},
	doi          = {10.3390/vehicles6040086},
	issn         = {2624-8921},
	url          = {https://www.mdpi.com/2624-8921/6/4/86}
}

@article{ribeiro2024supply,
	title        = {{SUPPLY: Sustainable Multi-UAV Performance-Aware Placement Algorithm for Flying Networks}},
	author       = {Ribeiro, Pedro and Coelho, André and Campos, Rui},
	year         = 2024,
	journal      = {IEEE Access},
	volume       = 12,
	number       = {},
	pages        = {159445--159461},
	doi          = {10.1109/ACCESS.2024.3488574}
}

@article{xiao2025vision,
	title        = {{Vision-Based Learning for Drones: A Survey}},
	author       = {Xiao, Jiaping and Zhang, Rangya and Zhang, Yuhang and Feroskhan, Mir},
	year         = 2025,
	journal      = {IEEE Transactions on Neural Networks and Learning Systems},
	volume       = 36,
	number       = 9,
	pages        = {15601--15621},
	doi          = {10.1109/TNNLS.2025.3564184}
}

@inproceedings{alla2024sound,
	title        = {{From Sound to Sight: Audio-Visual Fusion and Deep Learning for Drone Detection}},
	author       = {Alla, Ildi and Olou, Herv\'{e} B. and Loscri, Valeria and Levorato, Marco},
	year         = 2024,
	booktitle    = {Proceedings of the 17th ACM Conference on Security and Privacy in Wireless and Mobile Networks},
	series       = {WiSec '24},
	pages        = {123–133},
	doi          = {10.1145/3643833.3656133},
	isbn         = 9798400705823,
	url          = {https://doi.org/10.1145/3643833.3656133},
	numpages     = 11
}

@article{tian2025uavs,
title = {UAVs meet LLMs: Overviews and perspectives towards agentic low-altitude mobility},
journal = {Information Fusion},
volume = {122},
pages = {103158},
year = {2025},
issn = {1566-2535},
doi = {https://doi.org/10.1016/j.inffus.2025.103158},
url = {https://www.sciencedirect.com/science/article/pii/S1566253525002313},
author = {Yonglin Tian and Fei Lin and Yiduo Li and Tengchao Zhang and Qiyao Zhang and Xuan Fu and Jun Huang and Xingyuan Dai and Yutong Wang and Chunwei Tian and Bai Li and Yisheng Lv and Levente Kovács and Fei-Yue Wang},
}

@inproceedings{lin2024airvista,
	title        = {{AirVista: Empowering UAVs with 3D Spatial Reasoning Abilities Through a Multimodal Large Language Model Agent}},
	author       = {Lin, Fei and Tian, Yonglin and Wang, Yunzhe and Zhang, Tengchao and Zhang, Xinyuan and Wang, Fei-Yue},
	year         = 2024,
	booktitle    = {2024 IEEE 27th International Conference on Intelligent Transportation Systems (ITSC)},
	volume       = {},
	number       = {},
	pages        = {476--481},
	doi          = {10.1109/ITSC58415.2024.10919532}
}

@inproceedings{coelho2025a4fn,
	title        = {{A4FN: an Agentic AI Architecture for Autonomous Flying Networks}},
	author       = {Coelho, André and Ribeiro, Pedro and Fontes, Helder and Campos, Rui},
	year         = 2025,
	booktitle    = {2025 IEEE 36th International Symposium on Personal, Indoor and Mobile Radio Communications (PIMRC)},
	volume       = {},
	number       = {},
	pages        = {1--6},
	doi          = {10.1109/PIMRC62392.2025.11275590}
}

\end{document}